\documentstyle[twoside,fleqn,espcrc2]{article}
\def\ltap{\;\raisebox{-.5ex}{\rlap{$\sim$}} \raisebox{.5ex}{$<$}\;}

\newcommand{\be}{\begin{equation}}  
\newcommand{\ee}{\end{equation}}
\newcommand{\ba}{\begin{eqnarray}}  
\newcommand{\ea}{\end{eqnarray}}

\newcommand{\AmS}{{\protect\the\textfont2
  A\kern-.1667em\lower.5ex\hbox{M}\kern-.125emS}}

\title{Light Quenched Hadron Spectrum and Decay Constants on different
       Lattices}
\author{C.R.~Allton\address{Department of Physics, University of Wales Swansea, 
        Swansea, United Kingdom},
        V.~Gim\'enez\address{Dep. de Fisica Teorica and IFIC, Univ. de
        Val\`encia, E-46100 Burjassot (Val\`encia), Spain},
        L.~Giusti\address{Scuola Normale Superiore and INFN, Sezione di Pisa,
        Italy}\thanks{presented by L.~Giusti},
        F.~Rapuano\address{Dipartimento di Fisica, Universit\`a di Roma `La
        Sapienza' and INFN, Sezione di Roma, Italy}}
       
\begin{document}
\begin{abstract}
We present a study of ${\cal O}(2000)$ (quenched) lattice configurations from
the APE collaboration, for $6.0\le\beta\le 6.4$ using both the Wilson 
and the SW-Clover fermion action. We determine the light
hadronic spectrum and meson decay constants. 
We extract the inverse lattice spacing using data at 
the {\em simulated values} of the quark mass. We find an agreement with the 
experimental data of $\sim 5\%$ for mesonic masses and $\sim 10\%-15\%$ for baryonic 
masses and pseudoscalar decay constants. A larger deviation is present 
for the vector decay constants.    
\end{abstract}
\maketitle
\section{INTRODUCTION}
In the last four years the Ape group has been extensively studying lattice QCD
in the quenched approximation. Several simulations have been done to study weak 
matrix elements such as $f_D$, $f_B$, $B_K$ and to study semileptonic decays
\cite{latI}-\cite{latV}.
These simulations have allowed to study the dependence of the results on the 
spacing $a$ and to investigate finite size effects. Here we present results 
for light meson masses and decay constants 
and baryon spectroscopy. The results have
been obtained from eight sets of data at $\beta=6.0$, $6.2$ and $6.4$ using
either the Wilson action or the ``improved'' SW-Clover action, in order to reduce
$O(a)$ effects. The parameters used in each simulation 
are listed in Table~\ref{tab:latparams}. 
The Ape group has performed extensive comparisons on data 
extracted from smeared and non-smeared propagators and found no real
improvement for lattices with a time extent of 64 at $\beta=6.0$ and 
$\beta=6.2$ \cite{latII_a2}.
Note that the simulations have been performed at $\beta$ values of 
$6.0$ or larger. 
%tabella con le masse dei mesoni......
% space before first and after last column: 1.5pc
% space between columns: 3.0pc (twice the above)
\setlength{\tabcolsep}{.65pc}
%% -----------------------------------------------------
%% adapted from TeX book, p. 241
%\newlength{\digitwidth} \settowidth{\digitwidth}{\rm 0}
%\catcode`?=\active \def?{\kern\digitwidth}
%% -----------------------------------------------------
%\begin{tabular*}{\textwidth}{@{}l@{\extracolsep{\fill}}rrrr}
\begin{table}
\caption{Predicted meson masses in GeV for all lattices.}
\label{tab:mesonsGeV}
\begin{tabular}{llll}
\hline
& $M_\rho$ & $M_{\eta'}$ & $M_\phi$ \\
\hline
Exper.     & 0.770    & 0.686     & 1.019   \\
\hline
Lat~I   & 0.809(7) & 0.6849(3) & 0.977(7) \\
Lat~II  & 0.808(3) & 0.6849(1) & 0.978(3) \\
Lat~III & 0.81(1)  & 0.6849(5) & 0.98(1) \\
Lat~IV  & 0.803(6) & 0.6851(2) & 0.984(6) \\
Lat~V   & 0.79(1)  & 0.6856(5) & 1.00(1) \\
Lat~VI  & 0.797(7) & 0.6853(3) & 0.989(7)\\
Lat~VII & 0.796(4) & 0.6853(2) & 0.990(4)\\
Lat~VIII& 0.792(4) & 0.6855(2) & 0.994(4)\\
\hline
\end{tabular}
\end{table}
This is to negate
the large systematic error present in lattice data for $\beta \ltap
6.0$ due to lattice artifacts \cite{chris}. All the results we
obtained will be described in greater detail in a forthcoming paper \cite{lavoro}.
\section{MESON MASSES AND DECAY CONSTANTS}
Meson masses and decay constants have been extracted from two-point correlation
functions of the following local operators  
\[
P_5(x) = i\bar{q}(x)\gamma_5q(x), \;\; V_k(x)=\bar{q}(x)\gamma_k q(x)\; ,
\]
\[
A_\mu(x) = \bar{q}(x)\gamma_\mu\gamma_5q(x)
\]
in the standard APE way \cite{lavoro}. 
We fit the correlation functions of these operators to a 
single particle propagator with a $sinh$ in the case of
axial-pseudoscalar function and with a $cosh$ in other cases. 
\begin{table*}[hbt]
% space before first and after last column: 1.5pc
% space between columns: 3.0pc (twice the above)
\setlength{\tabcolsep}{.55pc}
% -----------------------------------------------------
% adapted from TeX book, p. 241
\newlength{\digitwidth} \settowidth{\digitwidth}{\rm 0}
\catcode`?=\active \def?{\kern\digitwidth}
% -----------------------------------------------------
%\begin{tabular*}{\textwidth}{@{}l@{\extracolsep{\fill}}rrrr}
\label{tab:latparams}
\caption{Summary of the parameters of the runs analyzed in this work and 
time windows used in the fits.} 
\begin{tabular}{lllllllll}
\hline         %123456789
&Lat~I&Lat~II&Lat~III&Lat~IV&Lat~V&Lat~VI&Lat~VII&Lat~VIII\\
\hline
Ref & \cite{latI}& \cite{latI,latII_a} & \cite{latIII} & \cite{latIII}
    & \cite{latV} & \cite{latII_a,latII_a2} & \cite{latIII} & \cite{latIII} \\
$\beta$&$6.0$&$6.0$&$6.2$ &$6.2$&$6.2$&$6.2$&$6.4$&$6.4$\\
Action & SW & Wil & SW & Wil& SW & Wil & Wil & SW \\
\# Confs&200&320&250&250&200&110&400&400\\
Volume&$18^3\times 64$&$18^3\times 64$&$24^3\times 64$& $24^3\times 64$ 
&$18^3\times 64$&$24^3\times 64$&$24^3\times 64$&$24^3\times 64$\\
\hline
$K$&  -   &  -   &0.14144&0.1510&    -   &  -   &0.1488&0.1400\\
   &0.1425&0.1530&0.14184&0.1515& 0.14144&0.1510&0.1492&0.1403\\
   &0.1432&0.1540&0.14224&0.1520& 0.14190&0.1520&0.1496&0.1406\\
   &0.1440&0.1550&0.14264&0.1526& 0.14244&0.1526&0.1500&0.1409\\
\hline
\multicolumn{9}{c}{Mesons with zero momentum} \\
$t_1 - t_2$ & 15-28 & 15-28 & 18-28 & 18-28 & 18-28 & 18-28 & 24-30 & 24-30 \\
\hline
\multicolumn{9}{c}{Baryons with zero momentum} \\
$t_1 - t_2$ & 12-21 & 12-21 & 18-28 & 18-28 & 18-28 & 18-28 & 22-28 & 22-28 \\
\hline
\end{tabular}
\end{table*}
The pseudoscalar decay constant $f_{PS}$ has been 
extracted by combining the fit of $\langle A_0P_5\rangle$ with the ratio 
$\langle A_0P_5 \rangle /\langle P_5P_5\rangle $. 
The errors have been estimated by a jacknife procedure, blocking the data in
groups of 10 configurations and we have checked that there are no relevant 
changes in the error estimate by blocking groups of
configurations of different size. We have fitted the correlation 
functions in time windows
reported in Table~\ref{tab:latparams}. 
The time fit intervals are chosen with the following criteria: we fix the 
lower limit
of the intervals as the one at which there is a stabilization of the 
effective mass 
and as the upper limit the furthest possible point before the error overwhelms
the signal. We discard the possibility of fitting in a restricted region 
where a plateau is present, as the definition of such a region is highly 
questionable \cite{fukugita}. 
For lattices with highest number of configurations, i.e. LatII, LatVII and 
LatVIII, we
confirm that higher statistics do not lead to a longer or better 
(relative to the statistical error) plateau
\cite{fukugita}.
%tabella con le costanti di decadimento......
% space before first and after last column: 1.5pc
% space between columns: 3.0pc (twice the above)
\setlength{\tabcolsep}{.3pc}
%% -----------------------------------------------------
%% adapted from TeX book, p. 241
%\newlength{\digitwidth} \settowidth{\digitwidth}{\rm 0}
%\catcode`?=\active \def?{\kern\digitwidth}
%% -----------------------------------------------------
%\begin{tabular*}{\textwidth}{@{}l@{\extracolsep{\fill}}rrrr}
\begin{table}
\caption{Extrapolated/interpolated meson decay constants}
\label{tab:decaysGeV}
\begin{tabular}{llllll}
\hline
&$\displaystyle\frac{f_\pi}{Z_A m_\rho}$& 
$\displaystyle\frac{1}{f_\rho Z_V}$&$\displaystyle\frac{f_K}{Z_A m_{K^*}}$ 
&$\displaystyle\frac{1}{f_{K^*} Z_V}$& \\
\hline
Lat~I   &   0.17(1)  &   0.42(3)  &   0.172(9)  &  0.39(2) & \\
Lat~II  &   0.25(1)  &   0.51(2)  &   0.239(8)  &  0.48(2) & \\
Lat~III &   0.16(1)  &   0.39(3)  &   0.164(9)  &  0.36(2) & \\
Lat~IV  &   0.21(1)  &   0.47(2)  &   0.214(8)  &  0.45(1) & \\
Lat~V   &   0.19(3)  &   0.30(4)  &   0.18(2)   &  0.30(3) & \\
Lat~VI  &   0.21(1)  &   0.49(3)  &   0.21(1)   &  0.46(2) & \\
Lat~VII &   0.23(2)  &   0.39(2)  &   0.22(1)   &  0.38(1) & \\
Lat~VIII&   0.19(1)  &   0.30(2)  &   0.18(1)   &  0.29(1) & \\
\hline
\end{tabular}
\end{table}
Once the hadronic correlation functions have been fitted and the lattice
masses and matrix elements extracted, 
we extract as much physics as possible from the 
``strange'' region so that the chiral extrapolation will be 
needed only in few cases. The method we use is outlined below:\\
{\bf -} We define the lattice planes for meson masses and decay constants
($M_V a$, $(M_{PS} a)^2$),  ($f_{PS}
a/Z_A$, $(M_{PS} a)^2$) and ($1/(f_{V} Z_V)$, $(M_{PS} a)^2$) where 
the subscripts $PS$ and $V$ stand for pseudoscalar and vector meson. 
We plot the Monte Carlo data for each kappa used
in the simulation on these planes;\\
{\bf -} On the vector meson plane ($M_V a$, $(M_{PS} a)^2$) 
we impose the physical ratios $M_{K^*}/M_{K}$, $M_{\rho}/M_{\pi}$ and  
find the values of $M_\pi a$, $M_\rho a$ (only one independent), 
$M_K a$, $M_{K^*} a$ (only one independent), $M_{\eta '}a$ and $M_\phi a$;\\
{\bf -} We now use the value of meson masses determined above to read off the
lattice meson decay constants, $(f_\pi a/Z_A)$, $(f_K a/Z_A)$, 
$(f_{\rho} Z_V)^{-1}$ and $(f_{K^*} Z_V)^{-1}$
from the corresponding $f_{PS}$ and $f_V$ planes. \\
This procedure to extract physical quantities only requires meson
masses and not unphysical quantities such as quark masses or $k$
values. It allows us to study the $strange$ physics and fix the lattice spacing 
directly in the region where data have been simulated 
without chiral extrapolation to zero quark mass. This approach 
therefore reduces the
errors on physical quantities induced by the chiral extrapolation. 
Using the values of $a^{-1}$ from $M_{K^*}$ we have obtained the physical
value in GeV of meson masses reported in Table~\ref{tab:mesonsGeV}.
Comparing the vector meson mass (in lattice units) 
from lattices LatIII, LatV and \cite{ukqcd93} we
infer that there is the possibility of some residual finite volume effects 
on the $18^3$ lattice at $\beta=6.2$ \cite{lavoro}. This problem may also 
be present 
in our $\beta=6.4$ data for which the physical volume is
the same as in Lat V. Further investigations at larger 
lattice sizes are necessary to make the situation clearer.
Turning to the continuum limit, any dependence of the meson 
spectrum on $a$ is small and difficult to interpret unambiguously at this
stage.
In Table~\ref{tab:decaysGeV} we report results for the meson decay
constants without including the renormalization constants $Z_V$ and $Z_A$.
For both the pseudoscalar and vector decay constants we notice a difference
between the Wilson and SW-Clover data. This is presumably due to 
the different renormalization constants and the smaller
$O(a)$ effects in the latter case. There may also be a small residual 
finite lattice spacing effect in the vector decay constant in the SW-Clover
data which needs further study. 
Overall our results agree with experimental 
data to $\sim 5\%$ for meson spectrum and to $\sim 10\%-15\%$ for the 
pseudoscalar decay constants. 
In our opinion the vector decay constant deserves a much more careful study 
at larger volume and $\beta$.
\section{BARYON MASSES}
Baryon masses have been extracted from two-point correlation
functions of the following local operators
\begin{eqnarray}
N & = &\epsilon_{abc}(u^aC\gamma_5 d^b)u^c\nonumber\\
\Delta_\mu & = & \epsilon_{abc}(u^aC\gamma_\mu u^b)u^c\nonumber
\end{eqnarray}  
in the standard way by fitting the two point correlation functions 
to a single particle propagator with an $exp$ function. The errors have been
estimated as in the meson case.
%tabella con le masse barioniche...
% space before first and after last column: 1.5pc
% space between columns: 3.0pc (twice the above)
\setlength{\tabcolsep}{.19pc}
%% -----------------------------------------------------
%% adapted from TeX book, p. 241
%\newlength{\digitwidth} \settowidth{\digitwidth}{\rm 0}
%\catcode`?=\active \def?{\kern\digitwidth}
%% -----------------------------------------------------
%\begin{tabular*}{\textwidth}{@{}l@{\extracolsep{\fill}}rrrr}
\begin{table}
\caption{Predicted baryon masses in GeV for all lattices.}
\label{tab:baryonsGeV}
\begin{tabular}{llllll}
\hline
& $M_N $ & $M_{\Lambda\Sigma}$ & $M_\Xi $& $M_\Delta $ & $M_\Omega $\\
\hline
Exper.  & 0.9389  & 1.135   & 1.3181  & 1.232   & 1.6724 \\
\hline
Lat~I   & 1.09(5) & 1.21(4) & 1.32(4) & 1.3(1)  & 1.60(9) \\   
Lat~II  & 1.19(5) & 1.29(4) & 1.40(4) & 1.46(7) & 1.71(4) \\   
Lat~III & 1.1(1)  & 1.22(8) & 1.34(7) &  -      &  -   \\   
Lat~IV  & 1.17(7) & 1.28(6) & 1.39(5) &  -      &  -   \\   
Lat~V   & 1.1(2)  & 1.2(2)  & 1.4(1)  & 1.6(3)  & 1.9(2) \\   
Lat~VI  & 1.2(1)  & 1.3(1)  & 1.40(9) & 1.50(9) & 1.72(5) \\   
Lat~VII & 1.21(9) & 1.32(8) & 1.43(6) & 1.4(2)  & 1.72(9) \\   
Lat~VIII& 1.2(1)  & 1.29(8) & 1.41(7) & 1.3(2)  & 1.7(1) \\   
\hline
\end{tabular}
\end{table}
We have used the value of meson masses to read off the
lattice baryon masses from the planes ($M_N a$, $(M_{PS} a)^2$) and
($M_\Delta a$, $(M_{PS} a)^2$). Using the same values of $a^{-1}$ used for
meson masses we have obtained the results reported in Table~\ref{tab:baryonsGeV}.
For baryons we find very good agreement with the old APE \cite{parisi} 
data, while we find slightly
larger values when comparing with JLQCD \cite{fukugita} and LANL \cite{gupta}. 
Also for baryon masses we can conclude that we do not see
a dependence on $a$ and that we have an agreement with the 
experimental data of $\sim 10\%-15\%$. 


\begin{thebibliography}{9}
\bibitem{latI} %no60 & 6424_c  & 6424_w
M.~Crisafulli et al.,
\newblock {Phys. Lett.~B}~{369} (1996)~325-334.

\bibitem{latII_a} %1864 & 2464
APE Collaboration~(C.R.~Allton et al.),
\newblock {Nucl. Phys.~B}~(Proc. Suppl.) {34}~(1994)~456-458.

\bibitem{latII_a2} %1864 & 2464
APE Collaboration~(C.R.~Allton et al.),
\newblock {Nucl. Phys.~B}~(Proc. Suppl.) {34}~(1994)~360-362.

\bibitem{latIII} %fd62_w & fd62_c
APE Collaboration,
\newblock {Work in progress}

\bibitem{latV} %no62
APE Collaboration~(C.R.~Allton et al.),
\newblock {Phys. Lett.~B~326}~(1994)~295-302.

\bibitem{chris}
C.R.~Allton,
\newblock {Nucl. Phys.~B}~{437}~(1995)~641.

\bibitem{lavoro} %nostro lavoro
C.R.~Allton et al., 
%V.~G\`imenez, L.~Giusti and F.~Rapuano,
\newblock {Preprint in preparation}.

\bibitem{fukugita}
JLQCD Collaboration~(S.~Aoki et al.),
\newblock {Nucl. Phys.~B}~(Proc. Suppl.)~{47} (1996)~354. %{hep-lat/9510013}

\bibitem{ukqcd93}
UKQCD Collaboration~(C.R.~Allton et al.),
\newblock Phys. Rev.~D~49~(1994)~474-485. %hep-lat/9309002

\bibitem{parisi}
APE Collaboration~(S.~Cabasino et al.),
\newblock Phys. Lett.~B~258(1991)~195-201.

\bibitem{gupta}
T.~Bhattacharya, R.~Gupta, G.~Kilcup and S.~Sharpe,
\newblock Phys. Rev.~D~53 (1996)~6486-6508.          %hep-lat/9512021
\end{thebibliography}
\end{document}